\documentclass{emulateapj}

\usepackage[colorlinks=true,linkcolor=blue,citecolor=blue]{hyperref}
\slugcomment{ApJ Main Journal; accepted: 20 May 2014}

\newcommand{\swift}{\textit{Swift}}

\newcommand{\WISE}{\textit{WISE}\ }

\def\OII{[O\,{\sc ii}]\,}

\shorttitle{The host galaxy of the short GRB 071227}
\shortauthors{Nicuesa Guelbenzu et al.}

\begin{document}


\author{
A.~Nicuesa Guelbenzu\altaffilmark{1}, 
S. Klose\altaffilmark{1}, 
M. J. Micha{\l}owski\altaffilmark{2}, 
S. Savaglio\altaffilmark{3,9}, 
D. A. Kann\altaffilmark{1}, 
A. Rossi\altaffilmark{1,4}, 
L. K. Hunt\altaffilmark{5}, \\
J. Gorosabel\altaffilmark{6,7,8},
J. Greiner\altaffilmark{3}, 
M. R. G. McKenzie\altaffilmark{2},
E. Palazzi\altaffilmark{4}, 
S. Schmidl\altaffilmark{1} 
}

\altaffiltext{1}{Th\"uringer Landessternwarte Tautenburg, Sternwarte 5, 
07778 Tautenburg, Germany}
\altaffiltext{2}{Scottish Universities Physics Alliance, Institute for 
Astronomy, University of Edinburgh, Royal Observatory, Edinburgh, EH9 3HJ, UK}
\altaffiltext{3}{Max-Planck-Institut f\"ur Extraterrestrische Physik, 
Giessenbachstra\ss{}e, 85748 Garching, Germany}
\altaffiltext{4}{INAF-IASF Bologna, Via Gobetti 101, I-40129 Bologna, Italy}
\altaffiltext{5}{INAF-Osservatorio Astrofisico di Arcetri, Largo E. Fermi 5, 
I-50125 Firenze, Italy}
\altaffiltext{6}{Instituto de Astrof\'{\i}sica de 
Andaluc\'{\i}a, Consejo Superior
de Investigaciones Cient\'{\i}ficas (IAA-CSIC), Glorieta de la
Astronom\'{\i}a s/n, E-18008, Granada, Spain}
\altaffiltext{7}{Unidad Asociada Grupo Ciencia Planetarias 
UPV/EHU-IAA/CSIC, Departamento
de F\'{\i}sica Aplicada I, E.T.S. Ingenier\'{\i}a, Universidad del
Pa\'{\i}s Vasco UPV/EHU, Alameda de Urquijo s/n, E-48013 Bilbao, Spain}
\altaffiltext{8}{Ikerbasque, 
Basque Foundation for Science, Alameda de Urquijo 36-5,
E-48008 Bilbao, Spain}
\altaffiltext{9}{Physics Department, University of Calabria, via P. Bucci, 
I-87036 Arcavacata di Rende, Italy}

\title{Another short-burst host galaxy with an optically obscured
high star formation rate: The case of GRB 071227}

\begin{abstract} 
We report on radio continuum observations  of the host galaxy of the short
gamma-ray burst 071227 ($z$=0.381)  with the Australia Telescope Compact Array
(ATCA). We detect the galaxy in the 5.5~GHz band with an integrated flux
density of $F_\nu = 43\pm11~\mu$Jy,  corresponding to an unobscured
star-formation rate (SFR)  of about 24~M$_\odot$/yr, forty times higher
than what was found from optical emission lines.  Among the $\sim$30
well-identified and studied host galaxies of short bursts this is the third
case where the host is found to undergo an episode of intense star formation.
This suggests that a fraction of all short-burst progenitors hosted in
star-forming galaxies could be  physically related to recent star formation
activity,  implying a relatively short merger time scale.
\end{abstract}


\section{Introduction \label{Intro}}

Gamma-Ray Bursts (GRBs) are divided into long/soft and short/hard,  with the
borderline at 2 seconds \citep{Kouveliotou1993}. Other more physically
motivated indicators to classify GRBs have been proposed (e.g.,
\citealt{Zhang2009.703}), but this division between long/soft and short/hard is
still a widely used classification.  It is commonly accepted that  long GRBs
are linked to the collapse of very massive stars 
(for a review, see \citealt{WB2006,Hjorth2012book}),  while short GRBs are
generally believed to originate from compact stellar mergers 
(for a review, see \citealt{Nakar2007,Berger2013arXiv1311.2603B}).  Recent
theoretical studies suggest that the population of short-burst progenitors is
dominated at low redshifts by merging neutron star (NS)-NS systems, while at
high redshifts merging black hole (BH)-NS binaries dominate the population
\citep{Dominik2013}.

Of particular interest is the question of the age of the merging compact
stars. Individual and qualitative age estimates of the progenitor can be
obtained if a host galaxy can be identified. Even though there is still a
substantial fraction of hostless short-burst afterglows
(\citealt{Tunnicliffe2014}), for more than 50\% of all well-localized events
host galaxies have been found. While in elliptical galaxies  at $z\lesssim1$
the GRB progenitors are most likely members of an old stellar population, in
star-forming galaxies they could also be young.  In fact, short-burst
progenitors are found to be  hosted by galaxies of all morphological types
(e.g., \citealt{Leibler2010,Fong2013,Fong2013ApJ...769...56F}) and stellar
evolution models  suggest a broad range in the ages of merging binaries,
reaching from tens of millions to billions of years \citep{Belczynski2006}.

Star-formation rates (SFRs) in GRB host galaxies  are usually derived from
measured emission line fluxes, e.g., \OII in case of  redshifts
$z\lesssim1$. These measurements suffer however from the  unknown  extinction
in these galaxies and, therefore, represent  only a lower limit on the present
SFR. On the other hand, radio observations are unaffected by extinction and
provide an unobscured view on the star-forming activity in a galaxy via
synchrotron radiation from electrons accelerated in supernova remnants, though
over a relatively broad time span of about 100 million years.  Consequently,
in recent years radio observations of the host galaxies  of {\it long GRBs}
have been undertaken with the goal to derive the unobscured  SFR based on the
measured radio continuum flux (\citealt{Berger2003ApJ588.Radio,
  Stanway2010,Hatsukade2012,Michalowski2012,PerleyPerley.Radio.2013}).  In
most cases the hosts of long bursts could not be  detected at radio
wavelengths, with typical limits ranging from $\lesssim$10~M$_\odot$/yr for
low-redshift hosts to $\lesssim$1000~M$_\odot$/yr  for high-redshift hosts,
only in $\sim10$ cases are SFRs found to be $\sim$100~M$_\odot$/yr.  Since in
most cases these galaxies have optically deduced star-formation rates of at
most few M$_\odot$/yr, radio observations might simply not  have gone deep
enough to detect them (still leaving room for a substantial  optically hidden
star formation rate).

In constrast to increasing attempts to explore the hosts of {\it long bursts}
in the radio band, no systematic radio investigation of the hosts of {\it
  short bursts} has yet been performed. The only exception is  a single deep
radio observation  (and detection) of the host of the short GRB 120804A at
$z$=1.3 (\citealt{Berger2013.765}). In all other cases,  derived SFRs are
based on optical emission lines. However, it is entirely possible  that the
hosts of several short bursts are also affected by internal extinction,
possibly resulting in a substantial underestimation of the derived SFRs.  Like
long GRBs, short-burst afterglows can, in principle, be used to derive host
extinction based on multi-color afterglow data (e.g.,
\citealt{Kann2011,Nicuesa2012}). Nevertheless,  such an approach is in any
case limited to the line of sight towards the GRB progenitor and is not
necessarily  representative for the global host-galaxy extinction.

The host of the short burst 071227 is well-suited as a radio target in this
respect.  While the afterglow of GRB 071227 is known only from a single
detection in the $R_C$ band, and no line-of-sight extinction $A_V$ could be
derived, optical broad-band photometry of its host galaxy suggests a high
internal $A_V$ (\citealt{Leibler2010}).  Also, the galaxy is relatively
close-by ($z=0.381\pm0.001$; \citealt{Davanzo2009}), hinting at its potential
detection in the radio band.

GRB 071227 triggered \swift/BAT and had a duration of $T_{90}$(15-350~keV)
$=1.8\pm0.4$~s (\citealt{Sakamoto7147,Sato7148}).  Its intense first spike was
followed by extended soft emission lasting  for $\sim$100~s
(\citealt{Sakamoto7156}).  The burst was also detected by \emph{Konus-Wind}
(\citealt{Golenetskii7155}; duration $\sim$1.7~s) and \emph{Suzaku-WAM}
(\citealt{Onda7158}; $T_{90}\sim$1.5~s). \swift/XRT localized a bright X-ray
afterglow (\citealt{Beardmore7153}); \swift/UVOT revealed a single faint
source near the XRT error circle, which was identified as a galaxy also
visible in the DSS (\citealt{Berger7151}).  Later observations revealed that
the galaxy is an edge-on, late-type galaxy (\citealt{Davanzo2009}).  The
optical afterglow was situated at a projected distance of about 15 kpc away
from the galactic nucleus, close to the galactic plane. 
Its SFR was determined via the observed \OII line flux to
be a modest $\sim$0.6~M$_\odot$/yr (\citealt{Davanzo2009}). 

Here we report on radio continuum observations of the GRB host galaxy with the
aim to measure its unobscured star-formation rate.  Throughout the paper, we
adopt a $\Lambda$CDM cosmology  with $\Omega_M=0.27$, $\Omega_{\Lambda}=0.73$,
and $H_0=71$~km/s/Mpc (\citealt{Spergel2003}). At the given redshift  of
$z$=0.381 a value of $1''$ corresponds to a projected distance of 5.2 kpc.  

\section{Observations and data reduction}

Radio continuum observations of the host of GRB 071227 were performed on 26/27
July 2013 in the 5.5 and 9.0~GHz  bands (corresponding to 6 and 3~cm) with the
Australia Telescope Compact Array  (ATCA) using the upgraded Compact Array
Broadband Backend (CABB) detector (\citealt{Wilson2011}) and all six 22-m
antennae with the 6~km baseline (program ID: C2840). In this configuration the
width of the primary beam is 9 (at 5.5~GHz)
and 5 arcmin (9.0~GHz), respectively (ATCA users
guide\footnote{\tt http://www.narrabri.atnf.csiro.au/observing/users$_-$guide/
  html/new$_-$atug.pdf}), and the synthesized beam sizes are
$\sim$2\farcs3$\,\times\,$2\farcs9 and $\sim$1\farcs4$\,\times\,$1\farcs8, at
6 and 3\,cm, respectively.  CABB integrates in both bands simultaneously with
2048 channels, beginning at 4.476 and 7.976 GHz, respectively, with an
increment of 1 MHz. Bandpass and flux calibration was performed by observing
PKS B1934--638 for 10 min at the beginning  of the observing run. Phase
calibration was done by observing the source 0308--611 for 3~min every hour
followed by 57 min integrations on target. Altogether 11 such 1-hr cycles were
executed, in total covering nearly a complete ($u,v$) plane. 

Data reduction was performed using MIRIAD version 1.5 for ATCA radio
interferometry.\footnote{\tt
  http://www.atnf.csiro.au/computing/software/miriad/} All data were cleaned
from radio frequency interference (RFI) using the {\tt pgflag}
routine. The effective bandwidth was thus reduced by 13.6\% to 1770~MHz.
Calibrated and RFI-cleaned visibilities were finally Fourier
transformed using the ``robust'' weighting  option, varying this parameter
between ``natural'' and ``uniform'' weighting and selecting the one that gave
the best result, reaching a rms value of 7$\mu$Jy at 5.5~GHz and $5\mu$Jy at
9.0~GHz.  Flux measurements were done using the {\tt imfit} task by fitting a
two-dimensional Gaussian profile.

\begin{figure}[t]
\includegraphics[width=8.6cm,angle=0]{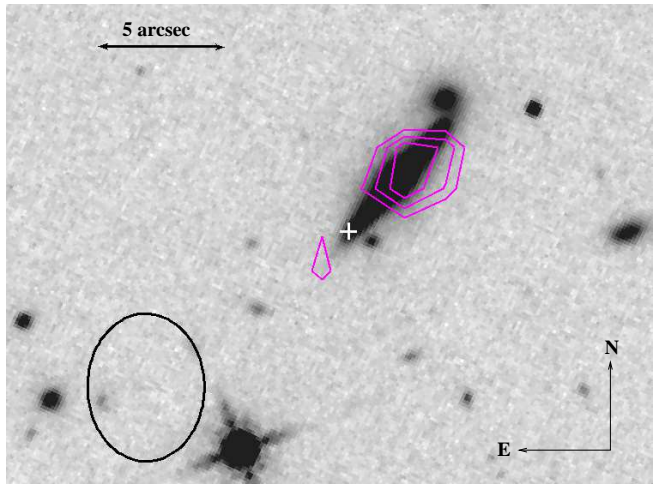}
\caption{
The host galaxy of GRB 071227 is detected with ATCA  in the 5.5~GHz band with
an integrated flux density of $F_\nu$ = 45~$\mu$Jy. The underlying optical
image (HST; \citealt{Fong2013})  shows the  host galaxy at $z$=0.381 as well
as the position of the optical afterglow (white cross).  Radio contour levels
(drawn in magenta) correspond to $F_\nu$= 10, 12.5, and 15$\mu$Jy
(image RMS = 4.8$\mu$Jy). The size of
the shown synthesized beam is 2\farcs3$\,\times\,$2\farcs9 (minor axis, major
axis).}
\label{071227}
\end{figure}

\section{Results}

We detect the GRB host galaxy in the 5.5~GHz band with a total integrated flux
density of $F_\nu = 43\pm11~\mu$Jy. The flux is nearly centered at the bulge
of the galaxy. The fact that the radio emission is not exactly centered at
the optical center of the bulge (offset $\sim$0\farcs4) could be an additional
interesting feature. However, given our synthesized beam size and
signal-to-noise ratio, the positional radio uncertainty is  0\farcs2 (as it
follows from the {\tt imfit} task under MIRIAD). Moreover,  the astrometric
uncertainty between the radio and the optical image has similar amplitude. The
0\farcs4 offset is therefore statistically not significant ($<3\sigma)$.
According to \cite{Davanzo2009}, during the spectroscopic observations the
slit was centered on the bulge of the galaxy with its orientation in
north-south direction. The \OII line flux reported by \cite{Davanzo2009}
therefore traces the region where we detect the  radio emission. 

No emission is detected from the position of the afterglow
(Fig.~\ref{071227}), although the galaxy is resolved as an extended object:
the deconvolved position angle of its radio image (--70.6$\pm$16.6 degrees)
agrees within its 1$\sigma$ error with the orientation of the galaxy in the
optical image. The galaxy is not detected in the 9~GHz band ($F_\nu <
26~\mu$Jy, 3$\sigma$); a limit on the spectral slope between both bands is
then about $-1.0\pm0.5$.

\begin{figure*}[t]
\includegraphics[width=18.0cm,angle=0]{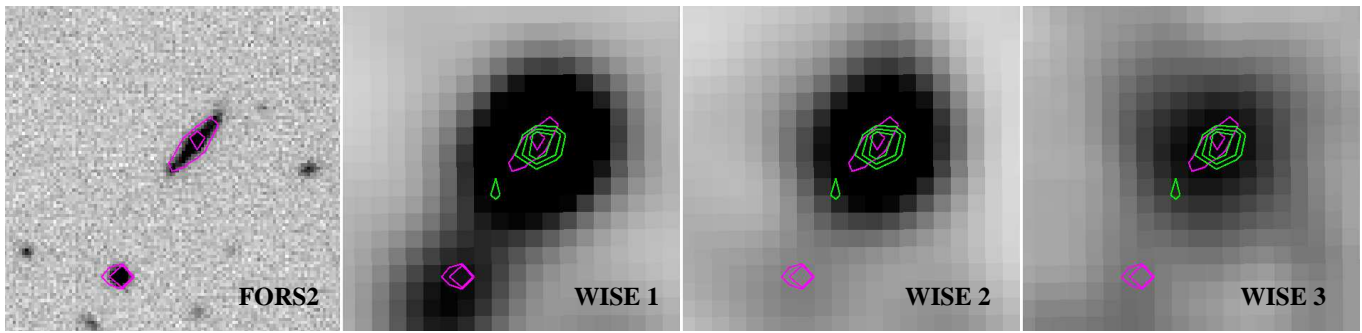}
\caption{
Multi-wavelength view of the host galaxy  covering optical, the \WISE 1
(3.4$\mu$m), \WISE 2 (4.6$\mu$m), and \WISE 3 (12$\mu$m) as well as the ATCA
5.5~GHz band. In green are shown the radio contours (Fig.~\ref{071227}), in
magenta the optical contour lines of the FORS2 image.}
\label{multiSED}
\end{figure*}

To further constrain the spectral energy distribution (SED) of the galaxy, we
explored the \WISE satellite data archive (\citealt{Wright2010}).  \WISE has
observed the entire sky in four bands at 3.4, 4.6, 12, and 22$\mu$m. The
catalog  lists only sources with a measured S/N greater than 5 in at least one
band.  \WISE detected the galaxy as a relatively bright source with AB
magnitudes of W1(3.4$\mu$m) = 18.27$\pm$0.03, W2(4.6$\mu$m) = 18.47$\pm$0.05,
W3(12$\mu$m) = 17.13$\pm$0.16, and W4(22$\mu$m) $>$16.2 (Fig.~\ref{multiSED}).
Combining these data with optical/NIR photometry from \cite{Leibler2010}, we
applied the SED fitting method detailed in
\citet{Michalowski2008,Michalowski2009,michalowski10smg,michalowski10smg4}
based on 35\,000 templates in the library of \citet{iglesias07}, plus some
templates of \citet{silva98} and \citet{Michalowski2008}, all developed in
{\sc Grasil} \citep{silva98}. They are based on numerical calculations of
radiative transfer within a galaxy, which is assumed to be a triaxial system
with diffuse dust and dense molecular clouds, in which stars are born. The
GRASIL fit\footnote{There are seven free parameters in the library of
\citet{iglesias07} (their table 3): the normalization of the Schmidt-type
law, the timescale of the mass infall, the intensity of the starburst, the
timescale of the molecular cloud destruction, the optical depth of molecular
clouds, the age of a galaxy and the inclination of a disk with respect to
the observer.  Parameter ranges used are, e.g., $A_V$=0 -- 5.5~mag and
metallicity from $4\,\times\,10^{-4}$ to 4.0 solar.} results in a SFR of
$\sim$24~M$_\odot$/yr, about 40 times higher compared to what  was derived
based on the \OII emission line flux. The mean host extinction  is
$A_V\sim$2.0 mag (see Sect.~\ref{sectab}), a value similar to what was deduced
by \cite{Leibler2010} based on their photometry and SED fitting. The infrared
data alone lead to  an estimated SFR of 40~M$_\odot$/yr, while using eqn. 6 in
\cite{Bell2003} the 5.5~GHz flux translates into an unobscured SFR of
$\sim$30~M$_\odot$/yr. These numbers characterize the host of GRB 071227 as a
galaxy that is undergoing an episode of intense star formation. 

\section{Discussion}

\subsection{Is GRB 071227 a member of the long-burst population?}

That the host of GRB 071227 is an actively star-forming  galaxy would not
attract  attention if the burst were a member of the long-burst population,
i.e., had a collapsar origin.  The short GRB 090426 (at $z$=2.609) is a good
example of a similar case in which the afterglow data strongly disfavor a
merger origin.  In the observer frame it had a duration of $T_{90}$ =
1.25$\pm$0.25~s (rest frame 0.33~s). However, several arguments have been put
forward that this burst was due to a collapsar event. These include the
afterglow luminosity, the circumburst density, the  spectral and energy
properties of the  GRB and the exceptionally high redshift for a short burst
(e.g., \citealt{NicuesaGuelbenzu2011a,Thoene2011MNRAS}, and references
therein).  Is GRB 071227 a similar outlier that contaminated the short-burst
data base?  In fact, at least its measured $T_{90}$ (Sect. 1) is, within the
error, in agreement with the classification as a long burst.

The position of the optical afterglow $\lesssim$ 0\farcs2 above the galactic
plane of its host galaxy ($\lesssim$ 1~kpc)  suggests that the progenitor of
GRB 071227 could have been a collapsing massive star inside a star-forming
region within the disk. However, no GRB-supernova (SN) was detected down to
$R_C=$ 24.9 (\citealt{Davanzo2009}), in spite of the favorable small redshift
of the host galaxy, implying a peak luminosity $L$(SN~071227) $<$ 0.1 $L($SN
1998bw)\footnote{SN 1998bw peaked at  $L_{\rm bol} \sim 1\,\times\,10^{43}$
  erg/s (\citealt{Pian2006Natur}) with $M_V \sim -19.2 + 5\log h_{65}$
  (\citealt{Galama1998})}.  Unfortunately, this non-detection  is not very
constraining for three reasons: (1) If this event had been seen across the
galactic plane of its host, then a host  extinction $A_V \gtrsim 2.5$ mag
could have dimmed the SN  light below the detection threshold.  (2) Given the
observed wide spread in GRB-SN luminosities (e.g.,
\citealt{Ferrero2006,Kann2011}),  a photometric detection of a SN following
this event was not expected with any certainty. (3) The long GRBs 060505 and
060614 did not even have any SN signal down to  hundreds of times fainter than
archetypal GRB supernovae (\citealt{Fynbo2006Natur}). 

A strong argument against a collapsar origin of GRB 071227 comes from the
properties of its prompt emission. Firstly, the spectral  lag\footnote{the
  arrival time difference between high and low-energy photons} of the first,
intense spike is consistent with zero (\citealt{Sakamoto7156}), a feature
found to be characteristic of short bursts
(\citealt{Norris2006ApJ,Zhang2006MNRAS.373}).  Secondly, in the Amati $E_{\rm
  iso} - E_{\rm peak}$ diagram (\citealt{Amati2008}), where long and short
bursts separate, the hard spike of GRB 071227 lies in the region occupied by
short bursts. \cite{Caito2010} have discussed in detail GRB 071227 and
concluded that the properties of its prompt emission define the burst as a
member of the short-burst population. Thirdly, the observed extended emission
of the burst (Sect. 1) is similar to  about ten other cases of short bursts
which can be understood as  magnetar-powered GRBs (\citealt{Gompertz2014}). In
addition, we also note that the luminosity of the optical afterglow of GRB
071227 (\citealt{Davanzo2009}) lies in the region occupied by short GRBs
(\citealt{Kann2011}).

\subsection{Comparison of the GRB 071227 host galaxy with other short-burst 
hosts \label{sectab} }

The high SFR puts the host of GRB 071227 in line  with the hosts of the short
GRBs 100206A (SFR$\sim$30~M$_\odot$/yr, $z$=0.407;
\citealt{Perley2012.758.GRB100206}) and 120804A (SFR$\sim$300~M$_\odot$/yr,
$z\sim$1.3; \citealt{Berger2013.765}). In particular, it is the third known
luminous infrared galaxy (LIRG) that hosted a short GRB. Among the 25
short-burst host galaxies compiled and  listed in
\cite{Berger2013arXiv1311.2603B} which have a measured SFR (bursts from mid
2005 to mid 2013), 19 have a SFR $\lesssim2.5~$M$_\odot$/yr (excluding here
GRB 071227) and three others lie between 6 and 16~M$_\odot$/yr.

\begin{figure}[t]
\includegraphics[width=8.2cm,angle=0]{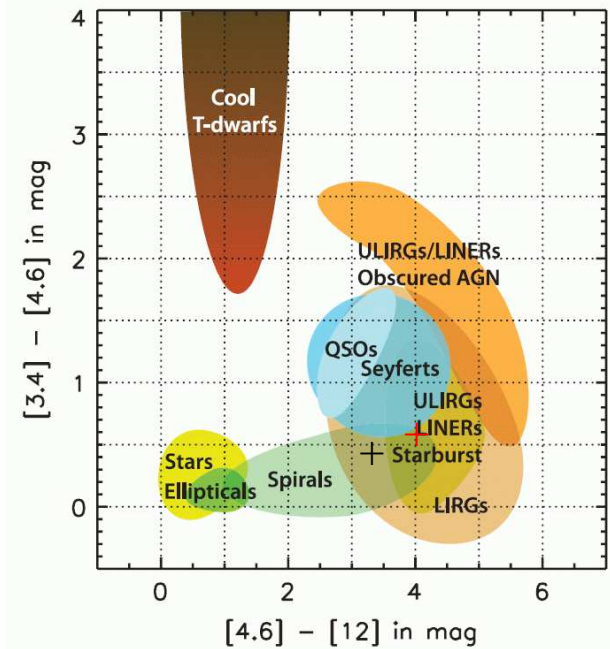}
\caption{
The position of the two starburst hosts of short bursts that have a \WISE
detection in the \WISE color-color diagram (Vega magnitudes, 
adapted from \citealt{Wright2010}), shown by crosses, in black GRB
071227, in red GRB 100206A. Both fall into the region occupied by
luminous infrared galaxies.}
\label{wise}
\end{figure}

\begin{figure*}[t]
\includegraphics[width=18.0cm,viewport= 10 643 454 850,clip]{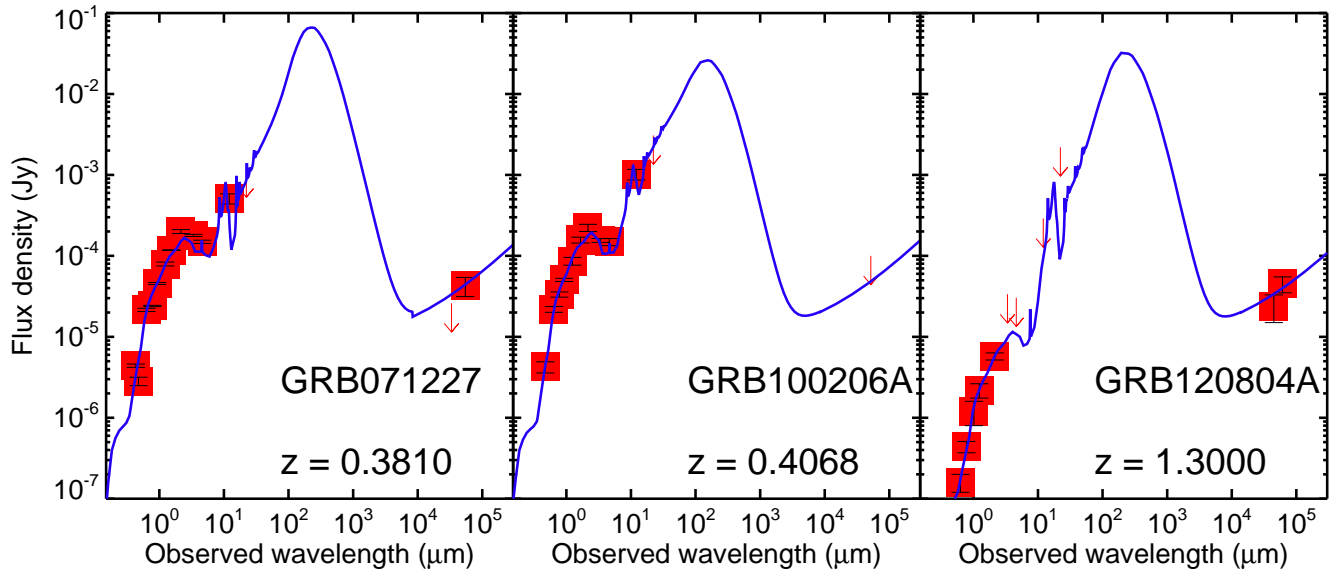}
\caption{
Spectral energy distribution of the three short GRB hosts with (U)LIRG SFRs
after correction for Galactic extinction.  {\it Red squares} are data points,
whereas {\it blue lines} denote {\sc GRASIL} models. For GRB 100206A and
120804A we made use of the  data published in \cite{Perley2012.758.GRB100206}
and \cite{Berger2013.765}, respectively. Note that the GRASIL fit assumes a
radio slope of $-0.75$. The host of GRB 071227 has a slightly steeper slope,
explaining the slight discrepancy between the 9~GHz upper limit and the fit.}
\label{071227.grasil}
\end{figure*}

In order to compare these galaxies with relatively high SFR, we collected  the
available multi-wavelength data of these galaxies  from the literature,
including the \WISE data, and performed GRASIL fits.  In
Fig.~\ref{071227.grasil} and Table~\ref{tab:GRASIL} we summarize the main
results of these fits.  We find that the hosts of GRB  071227 and 100206A have
similar IR luminosities and SFRs, while the corresponding values for the host
of GRB 120804A are ten times higher. All three hosts are fairly massive
dust-rich galaxies. In the \WISE color-color diagram (\citealt{Wright2010})
the first two galaxies fall into the region occupied by LIRGs
(Fig.~\ref{wise}).  The ultra-luminous host of GRB 120804A is not detected in
any \WISE band, likely due to its much higher redshift.

Are there more such infrared luminous galaxies that hosted a short burst? We
have compared the \WISE catalog (\citealt{Wright2010}) with the present sample
of short GRBs (as of January 2014) that have an error circle
$\lesssim$7~arcsec.\footnote{see the www pages maintained by J. Greiner and
D. A. Perley at http://www.mpe.mpg.de/jcg/grbgen.html and
http://www.astro.caltech.edu/grbox/grbox.php}  Results show that the hosts
of GRBs 071227 and 100206A stand out due to their high brightness in the W3
band. Among all short-burst host galaxies with a known redshift the host of
GRB 071227 is the second brightest in W3, exceeded only by the host of GRB
100206A. There is  only one more short-burst host that has detections in the
W1, W2 {\it and} W3 bands, namely the suspected host of GRB 060502B ($z$=0.287;
\citealt{Bloom2007}), though here the GRB-host association is not nearly as
secure as in the case of GRB 071227.

While a \WISE detection naturally favors low-$z$ galaxies, a comparison of
long-burst hosts with galaxies hosting short bursts is worthwhile. We found
that only one out of five $z<0.5$ {\it long-burst} host galaxies in the TOUGH
host galaxy survey (\citealt{Hjorth2012}) has detections in W1, W2 {\it and}
W3. The comparably high percentage of WISE-detected short-burst host galaxies
could be indicative that a fraction may have short merger timescacles.

\begin{deluxetable}{lrr rl}
\setlength{\tabcolsep}{1.2pt} 
\tabletypesize{\footnotesize}
\tablecaption{Main results from our GRASIL fits}
\tablehead{
\colhead{host:} &
\colhead{071227} &
\colhead{100206A} &
\colhead{120804A} &
\colhead{}}
\startdata
SFR   (M$_\odot$/yr)         &  23.9     & 29.7      & 311   \\
SFR$_{\rm IR}$ (M$_\odot$/yr) &  40.4     & 38.9      & 394  \\
SFR$_{\rm UV}$ (M$_\odot$/yr) &  0.40     & 0.43      & 0.58 \\
log (L$_{\rm IR}$/L$_\odot$)  &  11.37    & 11.35     & 12.36 \\
log (M$_\star$/M$_\odot$)     &  11.51    & 11.46     & 11.44  \\   
A$_V$ (mag)                  &  1.98     & 1.24      & 3.44    
\enddata
\label{tab:GRASIL}
\end{deluxetable}

\subsection{Is the host galaxy of GRB 071227 an AGN?}

\citet{Michalowski2008}  compiled evidence that long-GRB hosts are probably
not powered by AGNs. \cite{Berger2013.765} discussed in detail the
radio-detection of the host of  the short GRB 120804A and concluded that it
appears unlikely that this host harbors an AGN.  And what is the status of GRB
071227?

Our radio detection at 5.5~GHz corresponds to a specific luminosity of $L_\nu=
1.6\,\times\,10^{29}$ erg/s/Hz.   Assuming a spectral slope  in the radio band
of $-0.75$, at 1.4~GHz we expect a flux density of 150 $\mu$Jy. According to
\cite{Huynh2005AJ} and \cite{Ballantyne2009ApJ}, for such a flux density
roughly 25\% of sources are AGNs.

X-ray emission is one of the principal characteristics of AGN activity. As an
upper limit on the X-ray flux of the GRB host galaxy we can take the
non-detection of the afterglow by \swift \ at $t$= 314~ks with $F_\nu$(0.3-10
keV)$<1.3\,\times\, 10^{-19}$ erg/s/cm$^{2}$ (see the \swift \ light curve
repository; \citealt{Evans2007a,Evans2009}). For the given redshift this gives
$L_X< 6.5\,\times\,10^{42}$ erg/s, which is unfortunately not very
constraining.

However,  there two reasons that make us believe that our radio detection is
most likely not due to AGN activity: (1) The galaxy's colors in the WISE
diagram (Fig.~\ref{wise}) place it significantly far from the region occupied by
AGNs. (2) As stressed in Sect.~3, at 5.5~GHz the galaxy does not appear as a 
point source but is extended with the position angle oriented along the 
orientation of the galactic plane.

\subsection{Was GRB 071227 due to a young progenitor system?}

Our radio observations show that GRB 071227 originated in a  galaxy that is
undergoing an episode of intense star formation.  Also the short GRBs 100206A
and 120804A were hosted in galaxies with high SFRs of $\sim 30\,$M$_\odot$/yr
(\citealt{Perley2012.758.GRB100206}) and $\sim 300\,$M$_\odot$/yr
(\citealt{Berger2013.765}), respectively (see also Table~\ref{tab:GRASIL}).
In light of the high SFR found for the host galaxy of the short GRB 100206A,
\cite{Perley2012.758.GRB100206} discussed the question whether the GRB
progenitor was a member of an old or a young stellar population, in other
words could it have been physically related to the recent high SFR in its
host? 

In the case of GRB 071227, a link between the high SFR in the bulge of its
host galaxy and its position about 15 kpc ($\sim3''$) away  from the galactic
bulge in the galactic  disk would require a (projected) kick velocity of the
NS binary of about $150/t_8$ km/s, where $t_8$ is the age of the GRB
progenitor in units of $10^8$ years. For $t_8\sim1$, the typical range in ages
for starbursts traced by radio continuum observations,  the required kick
velocity is  relatively modest when compared with the kick velocities of some
Galactic neutron stars (e.g., \citealt{Tomsick2012}), and might even be small
compared to NS-NS/NS-BH systems. On the other hand, whether 150 km/s is a
relatively  high or even an unreasonable high value for a short-burst
progenitor is difficult to evaluate, at least form the observational point
of view. While the majority of all short bursts with
well-defined host galaxies (e.g., \citealt{Fong2010,Leibler2010,
Berger2013arXiv1311.2603B,Fong2013ApJ...769...56F,Fong2013}) seems to have
relatively low space velocities (because they exploded close to galaxies,
suggesting that these are their hosts), several aparently host-less bursts are
also known (e.g., \citealt{Tunnicliffe2014}). The latter could imply high
kick velocities, even though in these cases substantial stellar  ages of,
say, 10 billion years, could finally relax the requirement for a high kick
velocity.

Alternatively, as discussed by \cite{Davanzo2009}, the progenitor could have
been born in a star-forming region  in the galactic disk and exploded (merged)
within  even shorter time scales of 10$^6$ to 10$^7$ years. This scenario
avoids the need for a high kick velocity and for a kick velocity vector
that lies nearly parallel to the orientation of the galactic plane. A possible
causal connection to the high SFR in the central parts of the galaxy could
nevertheless still exist, given that at least in the local universe LIRGs show
that star-forming activity in the nuclear regions is usually spread out in a
galaxy on kpc scales (\citealt{Alonso-Herrero2013}). Our radio observations
then just pick out the radio-brightest central region of the galaxy.

\section{Summary and conclusions}

We detect the host galaxy of the short GRB 071227 at 5.5~GHz with a flux
density of $\sim45\mu$Jy, implying a SFR of about 30~M$_\odot$/yr.  This is
the second host galaxy of a short burst detected in the radio band  and the
third short-burst host found to be an infrared luminous galaxy.  Having now
found a third such case among the relatively small sample of about 30
well-studied short-burst hosts suggests that some observed stellar merger
events could be physically related to recent star-forming activity. A subgroup
of short-GRB progenitors could then merge within relatively short time scales,
contrary to what characterizes the short-GRB progenitor population hosted by
elliptical galaxies. 

Similar to what has been found for the type Ia supernovae rate (e.g.,
\citealt{Aubourg2008}), also the short-GRB rate in a galaxy might depend on
its recent star-formation rate as well as its total mass.  In fact, all three
galaxies we have discussed here are found to be very massive
(Table~\ref{tab:GRASIL}).  More systematic surveys of the hosts of
short GRBs in the radio and infrared bands are needed to address this issue. 

\begin{acknowledgements}

A.N.G. acknowledges support by M. Guelbenzu and by grant DFG Kl 766/16-1.
M.J.M. acknowledges the support of the Science and Technology Facilities
Council. A.R. acknowledges support by TLS Tautenburg and DFG Kl/766
13-2. D.A.K.  acknowledges support by TLS Tautenburg and MPE Garching.
L.K.H., E.P, and A.R. are grateful to support from PRIN-INAF 2012/13.
S. Schmidl acknowledges support by the Th\"uringer Ministerium f\"ur Bildung,
Wissenschaft und Kultur under FKZ 12010-514.  A.N.G. and S.K. thank  Catarina
Ubach \& Sarah Maddison, Swinburne University, Ivy Wong, CSIRO Sydney, and
Jamie Stevens, CSIRO Narrabri, for helpful discussions and observing guidance
as well as H. Meusinger and C. Rudolf,  Tautenburg, for useful discussions.
The Australia Telescope is funded by the Commonwealth of Australia for
operation as a National Facility managed by CSIRO. This publication makes use
of data products from the Wide-field Infrared Survey Explorer, which is a
joint project of the University of California, Los Angeles, and the Jet
Propulsion Laboratory/California Institute of Technology, funded by the
National Aeronautics and Space Administration. We thank the  anonymous referee
for very useful comments, which helped to improve the manuscript, and 
for a rapid reply.

\end{acknowledgements}


\bibliographystyle{aa}

\end{document}